\documentclass[12pt,preprint]{aastex}
\usepackage{color}
\definecolor{Blue}{rgb}{0.3,0.3,0.9}

\shorttitle{Detection of 3.3 \micron~aromatic hydrocarbon feature in N49 with AKARI}
\shortauthors{J. Y. Seok et al.}

\begin{document}

\title{Detection of 3.3 Micron Aromatic Feature in the Supernova Remnant N49 with AKARI}

\author{Ji Yeon Seok and Bon-Chul Koo}
\affil{Department of Physics and Astronomy, Seoul National University,
    Seoul 151-742, Korea}
\email{jyseok@astro.snu.ac.kr}
\and
\author{Takashi Onaka}
\affil{Department of Astronomy, Graduate School of Science,
University of Tokyo, Bunkyo-ku, Tokyo 113-0033, Japan}

\begin{abstract}

We present an infrared study of the supernova remnant (SNR) N49 in the Large Magellanic Cloud with the near-infrared (NIR; $2.5-5$ \micron) spectroscopic observations performed by $AKARI$. The observations were performed as a coarse spectral mapping to cover most of the bright region in the east, which enables us to compare the distribution of various line emission and to examine their correlation. We detect the 3.3 \micron~aromatic feature in the remnant, which is for the first time to report the presence of the 3.3 \micron~aromatic feature related to a SNR. In the line maps of H$_2$ 1-0 O(3), 3.3 \micron~feature, and Br$\alpha$, the distribution of the aromatic feature shows overall correlation with those of other emissions together with regional differences reflecting the local physical conditions. By comparison with other archival imaging data at different wavelengths, the association of the aromatic emission to other ionic/molecular emission is clarified. We examine archival $Spitzer$ IRS data of N49 and find signatures of other polycyclic aromatic hydrocarbon (PAH) features at 6.2, 7.7, and 11.3 \micron~corresponding to the 3.3 \micron~aromatic feature. Based on the band ratios of PAHs, we find that PAHs in N49 are not only dominantly neutral but also small in size. We discuss the origin of the PAH emission in N49 and conclude that the emission is either from PAHs that have survived the shock or PAHs in the preshock gas heated by radiative precursor.
\end{abstract}


\section{Introduction}

Over the past 20 years, various researches on polycyclic aromatic hydrocarbon (PAH) in interstellar medium (ISM) have been conducted both observationally and theoretically. PAHs are revealed as abundant, ubiquitous, and dominant in the ISM of galaxies \citep[and references therein]{tiel08}, and one of the main resources of infrared (IR) emission from near- to mid- IR wavebands. There have been a number of observations to show a variety of PAH features at different wavelengths. Several major emission features are usually detected in diverse objects such as the 3.3 \micron~C--H stretching band, the 6.2 and 7.7 \micron~C--C stretching bands, and the 8.6 and 11.3 \micron~C--H in- and out-of-plane bending bands. 
 
In PAH processing, interstellar shocks are known to play a crucial role. It is largely accepted that PAHs could be either the remaining dust condensation nuclei that escaped the grain growth process in asymptotic giant branch ejecta or the fragmentation of dust grains through shattering collision in fast interstellar shocks \citep[e.g.,][]{tiel08,jones96}. Also, interstellar shock waves have been considered as one of the main mechanisms to destroy PAH molecules. In spite of plentiful observations of PAH features, however, there are still unanswerable questions on the role of interstellar shocks in the evolution of PAHs. Particularly, how supernova shocks affect PAH molecules is barely explored. Detection of PAH features in supernova remnants (SNRs) is unexpectedly rare, recalling that the one of their formation process is related to interstellar shocks. 

There have been considerable literatures dealing with the survivability of PAH molecules in shocked environments. In particular, since PAH emission has been categorized as one of the IR emission components expected for SNRs based on the $Spitzer$ imaging data \citep{reach06}, substantial efforts have been made in order to search for observational evidence of PAH emission in SNRs. By using $Spitzer$ Infrared Spectrograph (IRS) observations, \citet{tappe06} firstly reported the detection of $15-20$ \micron~hump attributed to C--C--C bending modes of large PAHs ($\sim4000$ C-atoms) with the weakly detected 11.3 \micron~PAH feature in SNR N132D, which is located in the Large Magellanic Cloud (LMC). Based on the lack of PAH features at $6-9~\micron$ and the large ratio of the $15-20$ \micron~hump to the 11.3 \micron~feature, they interpret that small PAHs are rapidly destroyed by thermal sputtering in the supernova blast wave. Among the $Spitzer$ IRS spectra of several galactic SNRs, some in \citet{neu07} and \citet{hew09} also show the major PAH features with strong ionic and/or molecular lines, yet both authors have not mentioned the association of the features with the SNRs in the papers. In most cases that PAH emission is observed in SNRs, no convincing evidence for the PAH emission intrinsic to the SNRs has been reported. This lack of detection of PAH features in SNRs results from the difficulty of discrimination between PAH emission from SNRs and that from other back/foreground sources. 

In this respect, the LMC staying away from the galactic disk, could be the best place to observe PAH emission from SNRs. Moreover, since PAH molecules in shock regions are supposed to be able to survive only in those with dense clumps \citep{micel10}, SNRs interacting with dense circumstellar or interstellar material would be good candidates for detecting PAH emission. Thus, we have targeted the SNR N49, one of the SNRs interacting with ambient molecular clouds in the LMC, with $AKARI$ infrared space telescope \citep{mura07}. N49 (SNR 0525--66.1) is a middle-aged SNR \citep[$\sim 6600$ yr,][]{park03}, and the surrounding dense ambient medium including complex filamentary structures suggests that the progenitor of the SNR is a B-type star without a strong stellar wind \citep{shul83}. Thanks to its sufficient brightness at almost all wavelengths, this remnant has been studied by various observations at multi wavelengths; X-ray \citep{park03}, ultraviolet (UV) \citep{bla00}, optical \citep{vanc92,bil07}, IR \citep{willi06,seok08}, and radio \citep{dick98,dick95}. The observational results have shown not only complex filamentary morphology in both optical and IR wavebands, but also a complete shell-type structure in X-ray and radio. By CO observations \citep{ban97}, N49 is suggested to be interacting with its nearby molecular cloud at southeast, which is thought to be responsible for mostly asymmetric morphology and brightness enhancement toward southeast seen at all wavelengths. In this paper, we present the near-IR (NIR) spectra of N49 in the unprecedented wavelength range ($\sim2.5-5~\micron$) and report the first detection of an aromatic feature at 3.3 \micron~in SNRs. 

\section{Observations}

We performed spectroscopic observations of N49 (SNR 0525--66.1) using the Infrared Camera \citep[IRC;][]{onaka07} on board $AKARI$. These observations are part of the several $AKARI$ mission programs, and the details of the observations are listed in Table \ref{tbl-1}. The data were taken with a grism in the IRC NIR channel, NG ($2.5-5~\micron$), in the common mode of slit spectroscopy (Ns) during the post He phase (cooled by the onboard cryocooler). The dispersion (wavelength increment per pixel) is $\sim0.01$ \micron, and the slit width is 5\arcsec, which gives a spectral resolution of $R\sim100$ at 3 \micron~\citep{onaka07}. As a coarse spectral mapping, we made fourteen pointed observations to the SNR covering a bright wedge-shaped feature in the east as well as relatively IR faint region at west. In order to obtain a background spectrum, we also carried out two independent observations toward neighbor regions avoiding diffuse emission from the remnant. Figure \ref{fig1} shows the positions of the IRC NG/Ns slits superposed on the IRC N3 band ($2.7-3.8~\micron$) image that was taken in part of the $AKARI$ large-scale survey of the LMC {\citep[PI: T. Onaka,][]{ita08}. The total integration time is 396 s for each observation. 

The data were reduced by using the standard IDL package for the IRC spectroscopy data reduction \citep[toolkit version 20090211,][]{ohy07}. General preprocesses such as dark subtraction and linearity correction were applied during the pipeline. The extraction apertures of the final spectra are determined for areas observed with more than one slit positions, and finally five regions, named ``P1'' to ``P5'' from east to west, are chosen as marked in Figure \ref{fig1} (white rectangles). To extract each spectrum, we integrated 21 pixels along the slit for source spectra (except ``P5'') and 17 pixels for background spectra, corresponding to about 31\arcsec~and 25\arcsec, respectively. For the ``P5'' spectrum, we extracted it from a shorter aperture (13 pixels corresponding to 19\arcsec) in order to avoid low surface brightness regions. The center and the length of extraction were determined to maximize the signal-to-noize ratio (S/N), to avoid contamination from any point sources, and to average the resultant spectra in the overlapping region for the final spectra. The final spectra including the background spectrum are shown in Figure \ref{spec}. Since the background spectrum does not show any line or continuum emission, we did not subtract it from the final spectra to not degrade the S/N. 

\section{IR spectrum of N49}
\subsection{\textit{AKARI} IRC NG spectrum}

We examined the IRC NG spectra for the SNR N49, and the detected lines are given in Table \ref{tbl-2}. The MIR emission in N49 is dominated by ionic line emission \citep{willi06}. However, there is no strong ionic line within this IRC NG wavelength coverage, so that we could not detect any ionic lines except marginal detection of [Fe \textsc{ii}] $\lambda$ 4.889 \micron. The most prominent lines are hydrogen recombination lines such as Br$\alpha~\lambda$ 4.052 \micron~and Br$\beta~\lambda$ 2.626 \micron, which are seen at all spectra except spectrum P1 (Fig. \ref{spec}). Since P1 is extracted from just outside of the bright H$\alpha$ region, there is no strong recombination line emission but only an interesting feature at 3.3 \micron~(see next). Although H$_2$ molecular lines are not as distinct as the recombination lines, several rotational lines such as H$_2$ 1-0 O(3) and 0-0 S(11) lines are also clearly detected. 

For the lines detected in most spectra, we measured their intensities by fitting Gaussians with linear baselines. The intensities with uncertainties are given in Table \ref{tbl-2}. The uncertainties are the 1 $\sigma$ errors from the line fits, and the upper limits at 2 $\sigma$ level are given if the line emission is not apparent. Most lines in Table \ref{tbl-2} are not blended with others, yet there are two cases (i.e. H$_2$ 1-0 O(5), Pf$\delta$ at 3.235, 3.297 \micron~and Pf$\beta$, H$_2$ 0-0 S(9) at 4.654, 4.695 \micron) that two lines are located close to each other. For those cases, we applied two Gaussian fits to estimate their intensities together with either the peak wavelength or the full width at half maximum (FWHM) fixed if necessary. In addition, we fixed peak wavelengths or FWHMs for a few relatively weak lines. 

In the case of P1 to P3 spectra, we notice that the feature at 3.3 \micron~could largely originate from the well-known aromatic C-H stretching transition at 3.3 \micron rather than Pf$\delta$ at 3.297 \micron. In order to be attributable to Pf$\delta$ emission, its intensity is abnormally stronger than theoretical value ($\lesssim$ 10\% of Br$\alpha$ intensity in Case B) while the other Br$\beta$ and Pf lines are generally consistent with the expected values (see Table \ref{tbl-2}). Moreover, the intensities of the 3.3 \micron~feature do not vary with Br$\alpha$ intensities (see Fig. \ref{spec}), and the measured widths of the feature appear to be wider (FWHM $\sim0.04-0.05$ \micron) than those of other ionic/molecular lines (FWHM $\sim0.03$ \micron~on average) that corresponds to the instrumental width \citep{onaka07}. These widths are consistent with the typical width ($\sim$ 0.04 \micron) of the 3.3 \micron~aromatic emission \citep{vand04}. Hence, we consider that the 3.3 \micron~feature mainly arises from the aromatic hydrocarbon emission due to C--H stretching mode, which is the first time that we detect this aromatic feature associated with a SNR. Although this 3.3 \micron~feature does not require a polycyclic molecular structure, it is referred to as a 3.3 \micron~PAH feature en bloc hereafter.

There are several minor PAH features near the 3.3 \micron~feature such as the weak 3.4 \micron~band, or the broad plateau at $3.2-3.6$ \micron~\citep[e.g.,][]{tiel08}. In the IRC spectra, these minor PAH features are not obvious (Fig. \ref{spec}). We could only find features at $3.4-3.5$ \micron~range from P2 and P3, but they do not look similar to each other (i.e., have different peak positions), and there seems no such feature at P1. We measured their intensities which may be regarded as an upper limit of the minor 3.4 \micron~PAH feature intensity considering that other minor features and/or the underlying plateau might be included (Table \ref{tbl-2}). Despite of our ambiguous detection, the presence/absence of the 3.4 \micron~feature could be an interesting issue, because this feature can give a constraint on the band carriers, which is still controversial. There are two interpretations for this feature; overtones \citep[i.e., anharmonicity of hot bands;][]{alla, gebal} and the C--H stretching emission from aliphatic side groups attached to PAHs \citep[and references therein]{job96, tiel08}. Excited PAH bands of C--H stretch transition can be shifted to longer wavelengths due to anharmonicity. In that case, the 3.4 \micron~intensity increases if excitation gets higher. Besides, smaller PAHs tend to have a higher ratio of the 3.4 to 3.3 \micron~bands for the same internal excitation energy \citep{gebal}. Meanwhile, the presence of aliphatic bonds may delineate a balance between reactions with C, C${^+}$, and H producing aliphatic groups and photodissociation of these peripheral groups. When the UV radiation is strong, the dehydrogenation of PAHs severely proceeds so that the 3.4 \micron~emission becomes weak. However, \citet{tiel94} suggested that ion bombardment can form the aliphatic bonds by hydrogenation of an amorphitized graphite surface in interstellar shocks. Although its detection is tentative, the possible presence of the 3.4 \micron~feature at P2 and P3 but P1 may indicate that the PAHs inside the SNR are generally smaller with a strong UV field or that aliphatic bonds are more abundant due to shock processing compare to a preshock region. Further investigation with a higher sensitivity and spectral resolution would be needed to understand this point more clearly.

\subsection{Brightness distribution of PAH, H$\alpha$, and H$_2$ emission}

Although the 3.3 \micron~feature can be attributed to PAH emission, it is important to confirm whether the PAH emission is really associated with the SNR. For this purpose, maps of H$_2$, PAH, and Br$\alpha$ emissions are constructed from all spectra covering the PAH-bright area in the east (Fig. \ref{map}). The pixel size of the map is the same as the original pixel size of the slit (1.46$\arcsec\times1.46\arcsec$), and the intensity at each pixel is calculated by  resampling them with a $nearest$-$neighbour~interpolation$ method\footnote{The resampling uses a kernel extending a single input pixel. See https://www.astromatic.net/pubsvn/software/swarp/trunk/doc/swarp.pdf} and averaging the data values that fall on the pixel. Then, the line maps are smoothed with a three-pixel Gaussian. The resultant line maps reveal the distribution of 3.3 \micron~PAH emission together with Br$\alpha$ and H$_2$ 1-0 O(3). In order to remove the contamination by Pf$\delta$ line emission, the PAH map is subtracted by the scaled ($\times$0.1) Br$\alpha$ map. The morphology of PAH emission seemingly has association with those of other emission. The peak of PAH emission is found to be located at the tip of the bright wedge-shaped emission region, and the overall shape also shows enhanced emission along the wedge-shaped region. This strongly supports that the PAH emission is indeed related to the SNR.

All line maps in Figure \ref{map} show the enhanced emission at the eastern region. Previous observations at the other wavelengths similarly show the brightness enhancement, which is interpreted as the effect of interaction between the SNR and the nearby molecular cloud at the southeast \citep{ban97, bil07}. The morphology of the PAH emission, however, has similarities and differences in details with respect to those of H$_2$ O(3) and Br$\alpha$ emission. The eastern boundaries of both H$_2$ and PAH emission are indented with respect to the H$\alpha$ boundary (e.g., a gap at the east side between the two white crosses), while the Br$\alpha$ emission shows a good agreement with it (also see Fig. \ref{aat},$(a)$ and $(c)$). In particular, it is noticeable that both H$_2$ and PAH exceed the H$\alpha$ boundary around the tip of the wedge-shape region, which is at the center of the interacting molecular cloud (i.e. P1 and P2 in Fig. \ref{fig1}). The two clumps of the PAH emission in the south (white crosses in Fig. \ref{map}) could be seen at the coincident positions in the H$_2$ map. Meanwhile, the peak position of the PAH emission is not matched with that of the H$_2$ but rather closer to that of the Br$\alpha$. In addition, it is interesting that relatively weak PAH emission is observed where H$_2$ emission is the strongest.

Figure \ref{h2fit} compares the $average$ intensities of the representative emission lines in individual slits from P1 (east) to P4 (west). Since the intensities (listed in Table \ref{tbl-2}) are those averaged over the slits, they may not directly describe the spatial variation but quantitatively show their trend. The peak intensity of PAH emission appears at P2 (more eastward) unlike others, and the emission at P1 (outside the shock boundary) still has a comparable intensity to that of the PAH emission at P2. We could not see any tight correlations of the PAH emission with either H recombination or H$_2$ molecular line emission, but it is notable that the recombination line intensity rapidly increases over the boundary (P2) whereas the molecular line intensity is steady at the east (even inside the boundary) and becomes stronger at P3. 
 
Besides the emission intensities, we examined the variation of physical quantities such as the column densities, $N$(H$_2$) and the gas temperatures, $T_{gas}$ with respect to the four spectra, assuming local thermodynamic equilibrium (LTE) estimated from the intensities of H$_2$ molecular lines. H$_2$ line intensities can directly provide the column density, $N(J_u)$ of each upper state of the transitions, $J_u$. An extinction correction is not applied because the extinction toward N49 is negligible according to the measured $E(B-V)=0.37$ \citep{vanc92} and the extinction curve of the ``average'' LMC model \citep{wein01}. Performing one-temperature LTE fits to the observed H$_2$ transitional lines with a given ortho-to-para ratio (OPR = 3), it is found that $N$(H$_2$) varies from $\sim1$ to 5$\times10^{16}$ cm$^{-2}$ (Fig. \ref{h2fit}), while $T_{gas}$ steadily remains at $\sim2000$ K. The column densities we derived are much lower than the typical total column densities for SNRs \citep[$\sim 10^{20}$ cm$^{-2}$, e.g.,][]{hew09}. This is probably because only H$_2$ lines of high upper state transition are used for our estimation. In the case for SNRs, an $ankle$-$like$ $curve$ is representatively seen in the H$_2$ level populations \citep[e.g.,][]{shin10}, so that H$_2$ pure rotational lines with low upper states such as H$_2$ S(0)$-$S(2) transitional lines mainly determine the total column density. Then, what we derived is likely to be the column density of a hot component ($T_{gas}\sim2000$ K) within the SNR. Therefore, with the current observations, we cannot explore the relation between total H$_2$ and PAH.  

The morphological characteristics of PAH emission with respect to molecular and ionic gas would be reconfirmed and clarified by comparing to narrow filter images with a higher spatial resolution. As complement, we use preexisting data, the H$_2$ 1-0 S(0) at 2.12 \micron~taken with the $Anglo-Australian~Telescope~(AAT)$ \citep{dick95}, the $Spitzer$ IRAC 8 \micron~\citep{willi06}, and the $Hubble~Space~Telescope~(HST)$ H$\alpha$ image \citep{bil07}. The 2.12 \micron~and the H$\alpha$ images generally show consistent distribution with the H$_2$ O(3) and Br$\alpha$ maps in Figure \ref{map}, respectively. In Figure \ref{aat}, we can confirm the features described above. The eastern boundaries of the 2.12 \micron~and the PAH emission show a good agreement, and both emission are almost absent in the western part of the SNR where the optical filaments are comparably bright as seen in P5 spectrum (Fig. \ref{spec}). The peak of the PAH emission, however, shows better correspondence to that of H$\alpha$ emission rather than that of 2.12 \micron. It is certain that the PAH emission exists beyond the H$\alpha$ boundary in east. Thus, the other observations of N49 similarly show the same morphological characteristics as our $AKARI$ line maps. 

The characteristics of the PAH emission in N49 may be summarized as follows: (1) The overall morphology of the 3.3 \micron~PAH emission has the similarity to that of the H$_2$ emission in the sense that both are spatially confined. In contrast, the peak of the PAH emission is located rather closer to that of the Br$\alpha$ emission. (2) Toward the region with the brightest H$_2$ emission, however, the PAH emission is relatively faint. (3) There is PAH emission beyond the shock front in the east i.e., beyond the east boundary in optical, and its peak position is near the center of the interacting molecular cloud (Fig. \ref{fig1}). In addition, it is interesting that the distribution of PAHs shows a good agreement with the bright emission seen in the IRAC 8 \micron~image. A noticeable feature in the 8 \micron~image, not covered by our AKARI observation, is the bright emission beyond the eastern boundary. The nature of this feature is described in more detail in \S~3.3.

\subsection{\textit{Spitzer} IRS spectrum}

The detection of the PAH emission at 3.3 \micron~implies the possibility of other major PAH band emission such as 6.2, 7.7, and 11.3 \micron~features. We examined the $Spitzer$ IRS archival data for PAH emission \citep[AOR 6586112,][]{willi06}. The IRS slits are centered on the brightest tip of the wedge-shape region (Fig. \ref{aat}, $(b)$), and partially overlap with the IRC slits (P2 and P3). Retrieving the IRS low-resolution (SL: $5.2-14.5$ \micron) Post-Basic Calibrated Data (PBCD) from the $Spitzer$ archive, we have extracted spectra by using the standard SPICE package with a full aperture for an extended source (slit width: 3.7\arcsec, aperture length: 50.4\arcsec~(28 pixels)). We have applied background subtraction by using different order spectra. One of the benefits from the background subtraction is that the residual of the instrumental pattern (so called ``jail-bar'' pattern\footnote{See http://ssc.spitzer.caltech.edu/IRS/irsinstrumenthandbook/61/}) seen weakly in the PBCD data can be efficiently removed without much loss of S/N. The final IRS spectrum is shown in Figure \ref{irs}. It is almost the same as the one published by \citet{willi06}, but shows small differences probably due to background subtraction. The strong ionic lines such as [Fe II] and [Ne II] at $\lambda\lambda$ 5.32 \micron~and 12.83 \micron, are detected together with several H$_2$ transitional lines up to 0-0 S(2) $\lambda$ 12.30 \micron~(Fig. \ref{irs}, $(a)$). As in \citet{willi06}, a non-zero background level is seen, presumably because the background level varies. Taking a close look at the PAH bands in the spectra, we could identify several PAH features (Fig. \ref{irs}, $(b)-(d)$). PAH emission of C--H out-of-plane bending mode at 11.3 \micron~is clearly detected, and weak C--C stretching features at 6.2 \micron~and 7.7 \micron~with ambiguous 8.6 \micron~emission of C--H in-plane bending mode can be seen. We derive the intensities of the PAH band features by simple summations with a linear baseline fit over the entire order. The intensities of the 6.2, 7.7, and 11.3 \micron~PAH features are estimated as $1.0\pm0.5$, $2.0\pm0.4$, and $1.6\pm0.1$ in units of $10^{-16}$~W m$^{-2}$, respectively. The quoted errors include the calibration uncertainty and 1 $\sigma$ fluctuation in the baseline spectra.

For the clearly detected 11.3 \micron~PAH feature, we compare its spatial distribution to those of H$_2$ 0-0 S(2) $\lambda$ 12.3 \micron~and [Ne II] $\lambda$ 12.8 \micron~along the slit (Fig. \ref{prof}). For comparison, the spatial profile of H$\alpha$ emission extracted from the $HST$ image is also shown. Although a sharp jump in the H$\alpha$ intensity is not seen, a shock front might be located near the place where the H$\alpha$ intensity rapidly increases. A dashed line in Figure \ref{prof} can be regarded as the SNR boundary. The emission outside the shock front could originate from ambient gas heated by radiative precursors, and the hint of the radiative precursor can be found in H$\alpha$ emission in the zoomed profile at lower intensity level (Fig. \ref{prof}, $Bottom$). Beyond the SNR boundary, continuous decreases of 11.3 \micron~PAH and H$_2$ S(2) emission are seen up to the position around the pixel zero. The bump of the PAH emission at $-4$ pixel in the preshock area is more likely due to a background fluctuation (the bump is located outside of the southern boundary of Fig. \ref{aat} $(b)$). The profile of PAHs shows smooth decrease near the shock front like that of H$_2$, but shows better agreement with that of [Ne II] beyond the peak to the downstream. This is consistent with the morphological characteristics of PAH emission seen in the line maps. Similar distribution between the 11.3 \micron~PAH emission and [Ne II] has been observed toward several objects. For example, in the planetary nebula, BD +30 3639, \citet{matsu} detected the coexistence of PAHs and ionized gas, which has been attributed to a slow destruction of PAHs in the ionized region with the high electron density. Then, the good agreement between the 11.3 \micron~emission and [Ne II] in N49 might imply that destruction of small PAHs slowly proceeds in the shocked region (see more \S~4.2).

As mentioned in the previous section, the $Spitzer$ IRAC 8 \micron~image is quite similar to the 3.3 \micron~PAH emission image (Fig. \ref{aat}, $(b)$). In addition, the structure protruding from the bright tip at east is only seen in the 8 \micron~image among the IRAC images, which infers that the $7-9$ \micron~PAH bands possibly account for the feature. In the IRS spectra, [Ar II] line at 6.98 \micron~is the strongest emission line within the IRAC channel 4 band. However, it is necessary to consider other contributors such as H$_2$ 0-0 S(5), S(4) at 6.9, 8.0 \micron~as well as possible $7-9$ \micron~PAH emissions. In order to examine their relative contributions to the IRAC 8 \micron~band, we measure the intensities of H$_2$ S(5) and [Ar II] lines. Since they are resolvable enough to estimate their intensities by using two Gaussian fits, we can obtain $1.76\pm0.34$ and $3.78\pm0.36$ in unit of $10^{-16}$~W m$^{-2}$ for H$_2$ S(5) and [Ar II], respectively. Then, the intensity of H$_2$ emission together with the 7.7 \micron~PAH feature intensity ($3.76\pm0.52\times10^{-16}$ W m$^{-2}$ in total) seems to be comparable to that of [Ar II], so that the emission seen in the IRAC 8 \micron~image is dominated by not only the [Ar II] but also the H$_2$ lines with the $7-9$ \micron~PAH features. Therefore, the similarity between the 3.3 \micron~PAH emission image and the 8 \micron~image could arise from $7-9$ \micron~PAH emission with H$_2$ line emission, although we cannot rule out that the PAHs emitting the feature can coexist with the ionized gas. There is non-zero continuum level in the IRS spectrum in Figure 6 that we attributed to the residual of the background subtraction. The dust emission and synchrotron (or free-free) emission from the SNR should be negligible at this wavelength, so that even if the non-zero continuum is real, it should come from unrelated object. Therefore, we conclude that, although we cannot completely exclude the contamination from the general ISM, the good spatial correspondence is mainly due to the $7-9$ \micron~PAH emission in the IRAC 8 \micron~band.


\section{Discussion}

\subsection{Physical properties of PAHs in N49}

The PAH emission features are expected to reflect the local physical conditions. In particular, the relative strength variations among the PAH features at different wavelengths are mainly attributed to their different charge states \citep[e.g.,][]{bake01}. PAH emission from C--H modes such as 3.3 and 11.3 \micron~features becomes stronger in a neutral state, whereas others from C--C modes such as 6.2 and 7.7 \micron~features become stronger in an ionic state. Generally, comparisons of the intensities of 3.3 \micron~and 11.3 \micron, both normalized to that of 6.2 \micron, are regarded as an indicator for the degree of ionization of PAHs \citep[e.g.,][and references therein]{tiel08}. Similarly, the intensity ratios of 6.2 \micron~to 11.3 \micron~and 7.7 \micron~to 11.3 \micron~are often used as well. In our observations, the intensities of 3.3 \micron~bands cannot be directly compared with those of other PAH bands in the IRS spectra, but from the IRS spectra, we calculate band ratios among 6.2, 7.7, and 11.3 \micron~PAH emission; $I_{6.2}/I_{11.3}=0.63\pm0.31,~I_{7.7}/I_{11.3}=1.25\pm0.26$. 

Our ratios show unusually weak $6-8~\micron$ PAH features relative to a 11.3 \micron~feature in comparison with the median of the ratios from the star-forming galaxies \citep[$I_{7.7}/I_{11.3}\simeq3.6$ in][]{smith07} or those for Galactic diffuse emission \citep[$I_{7.7}/I_{11.3}\simeq2.0-3.3$ in][]{sakon04}. Such low PAH 7.7/11.3 ratios have been reported in some of elliptical galaxies, which is reasonably interpreted as a result of a larger fraction of neutral PAHs due to soft radiation field from evolved stars \citep[e.g., $I_{7.7}/I_{11.3}\simeq1-2$ in][]{kane08}. Similarly, our ratios can indicate that the PAHs in N49 are dominantly neutral even though the environmental cause is different. Figure \ref{ratio} is a diagram of PAH band ratios ($I_{6.2}/I_{11.3}$ versus $I_{7.7}/I_{11.3}$) from various objects in literatures, which shows that these PAH band ratios are linearly correlated regardless of object type. Note that the PAHs are mainly neutral in the lower left and mainly ionized in the upper right. Interestingly, it is found that our ratios also follow the universal linear correlation between the two ratios. As far as we know, N49 is the only SNR where both $I_{6.2}/I_{11.3}$ and $I_{7.7}/I_{11.3}$ ratios are obtained. In N132D, which is another SNR with PAH emission, only an upper limit of $I_{6.2}/I_{11.3}$ ratio could have been derived \citep{tappe06}, and is marked in Figure \ref{ratio}. This upper limit suggests that the ionization fraction of PAHs in N132D is not high, which is consistent with the theoretical expectation that PAHs can become neutral within a short time scale in the postshock layer \citep[see Figure 5 in ][]{micel10}. It is known that the charge states of PAHs are determined by the balance between the photoionization and the electron/ion collisions. The charge state is proportional to the quantity $\sim G_0\sqrt{T}/n_e$ \citep[for a review, see][]{tiel08}, where $G_0$ is the intensity of radiation field in units of the Habing field ($1.6 \times 10^{-6}$ W m$^{-2}$), $T$ is the gas temperature, and $n_e$ is the electron density. The ones with small $I_{6.2}/I_{11.3}$ ratio (or with neutral PAHs) represent a region where $G_0\sqrt{T}/n_e$ is small. For a few well-studied photodissociation regions (PDRs), \citet{gal08} have empirically interpreted the PAH band ratio with the above physical properties, $I_{6.2}/I_{11.3}\simeq [G_0/3040~n_e$(cm$^{-3})](T_{gas}/10^3$ K)$^{1/2}$+0.53 valid in the range $400\lesssim G_0/(n_e/$1 cm$^{-3})(T_{gas}/10^3$ K)$^{1/2}\lesssim4000$.

As we will discuss in the next section, the detected PAH emission is probably either from the PAHs in the shocked molecular gas and/or from the PAHs in the preshock gas heated by radiative precursor. In dense molecular clouds, PAHs usually exist in a neutral (and anionic) state. Even after they experience continuous (C-type) shock waves, neutral PAHs predominate in all charged states because photo-ionization is not significant in these circumstances \citep{flow03}. The PAH emission could be also from PAHs in preshock gas heated by radiative precursor. In N49, by modeling the UV and optical spectrum, \citet{vanc92} find that preshock densities of $20-940$ cm$^{-3}$ in the velocity range, 40 to 270 km s$^{-1}$, are required for the denser regions with radiative emission. The strength of the UV precursor in radiative shocks (30 km s$^{-1}\lesssim v_s\lesssim200$ km s$^{-1}$) is calculated that $G_0\approx1.6n_0 v_{s7}$, where $v_{s7}$ is the shock velocity per 100 km s$^{-1}$ and n$_0$ is the preshock hydrogen nucleus density \citep[and references therein]{mck87}. The coefficient varies with the shock velocity, and it is 1.6 when $v_s=100$ km s$^{-1}$. Then adopting a preshock density as 150 cm$^{-3}$ at $v_s=100$ km s$^{-1}$ under the requirement that $\rho_0v_s^2$ remains constant \citep{vanc92}, $G_0$ becomes 240. This UV radiation heats a column of 10$^{19}$ cm$^{-2}$ hydrogen nuclei to $\sim5000$ K and fully ionizes it \citep[e.g., the case of $n_0=100$ cm$^{-3}$ and $v_s=100$ km s$^{-1}$ in][]{allen}. In this radiative precursor region, the ratio $G_0\sqrt{T}/n_e$ will be small ($\sim110$), so that the PAHs will be neutral. If we extrapolate the empirical relation of \citet{gal08}, we obtain $I_{6.2}/I_{11.3}\simeq 0.53$ which is consistent with the observed ratio. 

Meanwhile, the observed band ratio can be compared with theoretical studies by \citet{drain01}, which show how the charge state and size of PAHs affect relative strengths of three band emission, 6.2 \micron/7.7 \micron~versus 11.3 \micron/7.7 \micron~(see Figure 16 in their paper). Our observed fluxes give $I_{6.2}/I_{7.7}\simeq 0.50\pm0.27$ and $I_{11.3}/I_{7.7}\simeq 0.80\pm0.15$ which are located near the line for neutral PAHs. More interestingly, our high $I_{6.2}/I_{7.7}$ ratio indicates that small PAHs are dominant in N49 although the error is large. The existence of small PAHs can also be verified by $I_{3.3}/I_{11.3}$ ratio since larger PAHs produce relatively a strong 11.3 \micron~feature while the 3.3 \micron~emission is enhanced by small PAHs \citep[e.g., ][]{schut}. It is not simple to directly derive the $I_{3.3}/I_{11.3}$ ratio from our data because the 3.3 \micron~and 11.3 \micron~fluxes are observed by different telescopes and positions. However, we make a rough estimation to obtain both the surface brightness at the peak and the integrated intensity with the full aperture (50.4\arcsec) along the IRS slit from the $AKARI$ 3.3 \micron~line map (Fig. \ref{map}) and the $Spitzer$ spectrum (see the slit position in Fig. \ref{aat}). The $I_{3.3}/I_{11.3}$ ratios of the peak brightness and the total intensity are measured as $\sim0.43$ and $\sim0.30$, respectively. These ratios are in a high end of the observed range (in general, $0.2-0.3$), which can be accounted for by small PAHs unless the radiation is hard enough to excite large PAHs \citep{mori}. For 100 km s$^{-1}$ radiative shocks, the UV radiation from the shock is mostly in Ly$\alpha$ photons \citep{mck87}, so it might be difficult to directly compare the shock radiation to the radiation from a central source. However, no detectable PAH feature at longer wavelengths easily produced by large PAHs (e.g., 16.2 or 17.4 \micron~features) is more likely to support a small contribution of large PAHs to the 3.3 \micron~feature \citep[see Figure 9 of][]{willi06}. Therefore, the observed $I_{6.2}/I_{7.7}$ and $I_{3.3}/I_{11.3}$ ratios indicate the presence of small PAHs, which is consistent with the PAH formation by fragmentation from larger carbonaceous grains but not with the preferential destruction of small PAHs in shocked gas.

\subsection{Origin of PAH emission}

PAH emission has been detected toward few shock-associated regions. Due to the limited observational evidence, it is poorly understood how shocks play a role in PAH processing. Generally, there are two sides to SNR shocks in terms of PAH processing, i.e., shocks can produce PAH molecules by shattering larger dust grains \citep[e.g.,][]{jones96}, but shocks can also destroy them by collisions with energetic particles in shocked gas \citep[e.g.,][]{micel10}. \citet{micel10b} show that the destruction mechanism in a hot gas ($T\gtrsim3\times10^4$ K) is dominated by electron collision for small/medium size PAHs ($50-200$ C-atom). They describe a PAH lifetime in a hot gas as $\tau_0=N_\mathrm{C}/Jn_{\mathrm{H/e}}$, where $N_\mathrm{C}$ is the initial PAH size, $J$ is the rate constant for electrons, nuclear and electronic interactions, and $n_\mathrm{{H/e}}$ is the hydrogen/electron density. At a fixed PAH size, a PAH lifetime can be determined by the gas density and temperature. In a rarefied gas with fast shocks ($v_{s} \gtrsim$ 300 km s$^{-1}$), no PAH formation is expected, and most PAHs are completely destroyed soon after being swept up by shocks. In particular, smaller PAHs are eroded more easily for $T\le10^6$ K. Nevertheless, if we see the emission from PAHs just swept up, the PAH emission should be confined to the shock front. In a dense medium, however, as shocks become slower ($v_{s} \lesssim100-150$ km s$^{-1}$), PAHs might survive and/or be formed in those shocked environments \citep{micel10}. Then, the detection of PAH emission in the shocked regions requires either efficient formation of PAHs or shielded condition against complete destruction. Currently, observational difference between newly formed and survived PAHs is not obvious, but it is supposed that the newly formed PAHs undergo the same processing as the survived ones soon after. Meanwhile, PAH emission could be also produced outside the shocked region by radiative precursors of shocks. Hence, it is necessary to separately examine the dominant mechanism for PAH emission depending on the local shock condition.

The complex structures of N49 imply that physical conditions inside the remnant should significantly vary in localities. This remnant can be largely divided up into three kinds of environments related to the characteristics of the medium from which different emission arises. The hot gas with X-ray emission is generated by the fast shock propagating in the diffuse intercloud medium \citep[preshock density $n_0=0.9$ cm$^{-3}$, $n_e=27-2300$ cm$^{-3}$, $T\gtrsim7\times10^6$ K, and $v_s\sim730$ km s$^{-1}$ in][]{park03}. The dense ambient medium produces the bright optical filaments (i.e., ionic line emission) by shocks of $v\leq140$ km s$^{-1}$ \citep[$n_0=20-940$ cm$^{-3}$, $T\sim10^4$ K in][]{vanc92}. The electron densities of the filaments vary from $n_e\simeq70$ to 515 cm$^{-3}$, and the densest regions have the densities of $1000-1800$ cm$^{-3}$ \citep{bil07,vanc92}. Lastly, the molecular line emission such as H$_2$ lines can arise from the dense clumps where slow (nondissociative) shocks are impinging on \citep[in general, $n_0\gtrsim30$ cm$^{-3}$, $v_s\leq50$ km s$^{-1}$ in][]{drain93}. In N49, both shocked molecular and ionic gases interestingly show analogous distributions, and the PAH emission is likely to be associated with both gases. This is not surprising in the sense that N49 is known to be interacting with a nearby molecular cloud \citep{ban97}. The coexistence of both molecular and ionic shocks in a single SNR is occasionally seen in SNRs interacting with molecular clouds, since the shocked dense clumps emitting H$_2$ transitional line emission are immersed in a less dense medium that emits various ionic line emission after swept up by shocks \citep[e.g.,][]{chev99}. 

In the hot diffuse medium of N49 ($n_e=27$ cm$^{-3}$, $T=7\times10^6$ K), the lifetime of PAHs with N$_\mathrm{C}=200$ is $\sim$ 1 month by adopting the analytical fits to the rate constant \citep[$J\sim2.7\times10^{-6}$ cm$^3$ s$^{-1}$; Table 2 in][]{micel10b}. Even for large-sized PAHs or PAH clusters, the lifetime is not long enough to be detected \citep[e.g., $\sim$ 20 months for N$_\mathrm{C}=1000$, Fig. 8 in][]{micel10b}. To explain the observed PAH emission, protective environments in N49 are necessary, which have already been noticed in optical spectroscopic observations \citep{vanc92,bil07}. By using the echelle observations with the $HST$ image \citep[see Fig. 3 in][]{bil07}, some broad emission features in the echellegrams reveal a ``head-tail'' structure, of which the bright emission at the head has a smaller radial velocity offset from the systemic velocity than the fainter emission in the tail. This type of the structure could originate from shocks encountering the dense gas. While a shock is propagating through the preshock gas in the inner part of the dense bright regions, the shock velocity drops so that the gas deep inside the dense filament can avoid to be swept up by shocks, staying unshocked. The observed H$\alpha$ recombination lines at the systemic velocity in the optical spectrum support this.

Based on the above discussion, we propose a schematic model of the PAH emitting condition in N49 in Figure \ref{schem}. As a shock is retarded by a dense gas (Region II and shocked Region III), the temperature of the postshock gas can substantially decrease ($T_{\mathrm{gas}}\propto v_s^2$). The PAH lifetime under these circumstances can significantly increase because the rate constant rapidly declines below 10$^5$ K \citep{micel10b}. \citet{vanc92} have measured electron densities and electron temperatures for the bright optical filament of N49 as $\sim1\times10^3$ cm$^{-3}$ and $\sim1\times10^4$ K, respectively. These quantities lead to the lifetime of $1.2\times10^6$ yr for 200 C-atom PAHs (c.f., $J\sim5.2\times10^{-15}$ cm$^3$ s$^{-1}$), which is much longer than the age of the SNR. PAHs in the central region (unshocked Region III) of the molecular clumps (and the dense filaments) can be heated by UV photons that mainly originate from the radiative shell, more precisely in this case, the knotty optical filaments around the clumps. In the case of the PAH emission observed outside the shock front (seen eastward at the tip in Fig. \ref{map}), it is most likely that the preexisting PAH molecules are heated by the radiative precursors and produce the emission features. When the shock velocity is greater than 80 km s$^{-1}$, the ionizing UV fluxes produced by the precursor become effective \citep{shul79}. The bulk range of the shock velocity in N49 is measured as $\lesssim$ 140 km s$^{-1}$ \citep{vanc92}, and we indeed see the hint of the radiative precursor in the zoomed H$\alpha$ line profile (Fig. \ref{prof}). In summary, PAH emission can be associated with shocked ionic gas (Region II) or (shocked) molecular gas (Region III) when the shock is sufficiently retarded with existence of heating source such as UV photons (Fig. \ref{schem}). Also, as a shock is propagating and retarded inside the dense region, shocked ionic and molecular gas can hierarchically exist together with PAH emission. Then, the morphological correlation of the PAH emission to other emission would depend on the shock velocity and the preshock density.

The absence of PAH emission at the peak of the H$_2$ emission is interesting. In Figure \ref{aat}, the region with the brightest H$_2$ emission seems to coincide with one of the bright H$\alpha$ filaments, and it is found that the filament particularly shows the most extreme difference among H$\alpha$, [S II], and [O III] emissions in optical; the [O III] emission is absent while the [S II] emission is relatively stronger than H$\alpha$ \citep[see Fig. 1 in][]{bil07, vanc92}. According to shock models in \citet{vanc92}, [O III] emission is generally produced by faster shocks and cannot be produced for shocks with velocities $\le80$ km s$^{-1}$. Meanwhile, as the shock velocity decreases, the intensities of [S II] show an increasing tendency with respect to that of H$\beta$ in the model calculation. In this context, it is likely that relatively slow shocks ($\le80$ km s$^{-1}$) have reached the filament and have resulted in these differences in the optical spectra. If so, the faint PAH emission can be explained, because the shocks with $v_s\le80$ km s$^{-1}$ cannot produce sufficient UV radiation \citep{shul79} at the outer layer of the filament which is necessary to heat PAH molecules. For H$_2$ emission, however, hydrogen molecules can be collisionally excited by shocks with those velocity ranges, and produce bright emission. For $v_s\gtrsim25-30$ km s$^{-1}$, the line intensity is sensitive to the preshock density, not to the shock velocity \citep{burton}. Hence, even if PAHs can survive, they might not generate detectable emission due to the lack of heating sources, while it is possible to produce bright H$_2$ emission in the same environment. This may also explain the IRS spectra of the Galactic SNRs interacting with molecular clouds, where H$_2$ emission is prominent but the PAH emission is absent \citep[e.g.,][]{hew09}.


The detection of 3.3 \micron~PAH band feature from N49 is surprising in a sense that only small PAHs with $<100$ C-atoms or a size $<6$ \AA~can notably produce the feature \citep{drain07} while small PAHs can be easily destroyed by strong shocks. Even elliptical galaxies showing low PAH 7.7/11.3 \micron~ratios (i.e., mainly neutral PAHs) seem to have no significant 3.3 \micron~PAH emission due to the lack of small PAHs \citep{kane07}. Meanwhile, the comparison between our PAH band ratios ($I_{6.2}/I_{7.7}$ and $I_{3.3}/I_{11.3}$) and the numerical studies \citep{drain01, mori} support that small PAHs are dominant in N49. In addition, there is no indication of the $15-20$ \micron~feature in the IRS spectrum by \citet{willi06} (see their Fig. 9), which is a signature of large PAHs. In the case of N132D, however, the relatively strong $15-20$ \micron~hump compared to $6-11.3$ \micron~PAH features suggests that the large PAHs are dominant, which was interpreted as an evidence for the survival of large PAHs behind fast shock \citep{tappe06}. Thus, PAH processing in N49 and N132D is likely to differ from each other. This difference in PAH sizes can be explained by the different environments and evolutionary stages of the two SNRs: N49 is a middle-aged SNR \citep[$\sim6600$ yr,][]{park03} and is interacting with a nearby molecular cloud with a radiative shock ($v_s\simeq100$ km s$^{-1}$), while N132D is relatively young \citep[$\sim2500$ yr,][]{morse} and still has a fast shock in less dense environment \citep[$\bar{v}_s\simeq800$ km s$^{-1}$,][]{morse}. In a dense gas with a moderately low temperature ($\lesssim3\times10^4$ K), PAHs would survive regardless of the size. This is because the nuclear interaction with helium is the dominant process of destruction in that situation, which does not significantly depend on the size of PAHs \citep[see Fig. 5 in][]{micel10b}. Hence, small PAHs are $NOT$ preferentially destroyed in N49, which can reasonably explain our detection of the 3.3 \micron~PAH feature in the postshock region with the relatively low temperature. This might imply that a wide range of PAH properties can be found in SNRs.

Based on our observations, it might be difficult to understand in detail, for example, what  is a dominant heating mechanism for emitting PAH features in the SNR, and whether the PAHs are newly formed or not. Nevertheless, it seems obvious that one of the essential conditions for the survival/(re)formation of PAHs is the existence of dense ambient medium around a SNR which directly influences the evolution of shocks such as the shock velocity. In addition to this, the sufficient heating source such as UV radiation is necessary for survived PAHs to produce observable PAH emission from a SNR.

\section{Summary}    

We have carried out an infrared (IR) spectroscopic study on the supernova remnant, N49 in the Large Magellanic Cloud by using $AKARI$ IRC NG observations which cover most of the bright eastern regions including distinct filaments. Since the observations have been performed as a coarse spectral mapping, we are able to make spectral line maps and have compared the distribution of different emission features. In the $AKARI$ IRC spectra ($2.5-5~\micron$), we detect 3.3 \micron~polycyclic aromatic hydrocarbon (PAH) band features with several strong hydrogen recombination lines and moderate H$_2$ molecular lines. {\it To our knowledge, this is the first time that we observe the presence of the 3.3 \micron~PAH feature related to a SNR.} Our main results are summarized below: 

1.  The 3.3 \micron~PAH feature in the spectra are clearly distinguished from other shocked ionic/molecular lines in terms of the line width and the intensity variation with position. In the $AKARI$ line maps, the distribution of the 3.3 \micron~PAH emission shows overall association with those of other emissions such as Br$\alpha$ and H$_2$ 1-0 O(3), which indicates that the PAH emission indeed originates from the SNR and is associated with both ionic and molecular emission. In addition, there are morphological dissimilarities among the line maps in a local scale reflecting the different physical conditions such as shock velocity, and preshock density in each region. The morphological characteristics are also clarified and confirmed by comparison to archival $Spitzer$ 8 \micron, $HST$ H$\alpha$, and H$_2$ 2.12 \micron~images obtained at $AAT$. 

2.  Overall distribution of the 3.3 \micron~PAH emission is more similar to that of the H$_2$ emission in a sense that the H$_2$ emission is spatially confined but the Br$\alpha$ emission is extended over a large area. In addition, the 3.3 \micron~PAH emission is extended beyond the eastern shock boundary with the intensity comparable to those in the inner regions. These indicate that the PAH emission is possibly associated with the interacting molecular cloud, the center of which nearly coincides with the bright PAH emission region. Meanwhile, the peak position of the PAH emission is not matched with that of the H$_2$ emission but rather close to that of the Br$\alpha$ emission. In addition, the PAH emission is relatively faint at the peak of the H$_2$ emission.  

3.  We find signatures of other PAH features, C-C stretching modes at 6.2 and 7.7 \micron~and C-H out-of-plain bending mode at 11.3 \micron~in the archival $Spitzer$ IRS SL spectra ($5.2-14.5~\micron$). We derive the band ratios of PAHs, $I_{6.2}/I_{11.3}$ and $I_{7.7}/I_{11.3}$ from the IRS spectra ($I_{6.2}/I_{11.3}=0.63\pm0.31,~I_{7.7}/I_{11.3}=1.25\pm0.26$), which implies that the PAHs in N49 are dominantly neutral. This is consistent with the theoretical expectation for shocked PAHs in dense molecular clouds. It is found that the ratios follow the universal linear correlation between the two ratios in literatures. We try to associate the $I_{6.2}/I_{11.3}$ ratio with the physical quantity $G_0\sqrt{T}/n_e$ by using the empirical relation in \citet{gal08} under the assumption that the shock radiation is the dominant heating mechanism of PAHs in a very dense clump. In addition, the relatively high $I_{6.2}/I_{7.7}$ and $I_{3.3}/I_{11.3}$ ratios ($0.50\pm0.27$ and $\sim0.36$, respectively) indicate the existence of small PAHs according to the numerical studies \citep{drain01,mori}. These results are consistent with the PAH formation by fragmentation from larger carbonaceous grains but not with the preferential destruction of small PAHs in shocked gas.

4.  The morphological features of the PAH emission can be attributed to the different mechanisms of SNR shocks in terms of PAH processing. For the PAH emission associated with either shocked H$_2$ gas or (shocked) ionic gas, or both, PAHs must exist in dense gas where the shocks have been sufficiently retarded (even terminated) to avoid the complete destruction. Depending on the shock velocity and the preshock density, the PAH emission can be associated with either the ionic gas or the molecular gas. Although PAHs can survive a slow shock, detectable PAH emission may not arise due to the lack of UV radiation in a certain condition. For the PAH emission outside the SNR, the radiative precursor could be responsible for the excitation. For PAHs to exist and radiate in SNRs, an ambient dense medium and a sufficient heating source around the medium are most likely to be required.

\acknowledgments
This work is based on observations with AKARI, a JAXA project with the participation of ESA. We wish to thank all the members of the AKARI project. We would like to thank John Dickel for providing the $AAT$ data and E. R. Micelotta for helpful discussions about PAH processing in a shocked region. We are also grateful to J.-H. Shinn for useful interactions about the calculations of H$_2$ level population. This research was supported by Basic Science Research Program through the National Research Foundation of Korea
(NRF) funded by the Ministry of Education, Science and Technology
(NRF-2011-0007223).

\clearpage

\begin{figure}
\epsscale{0.9}
\plotone{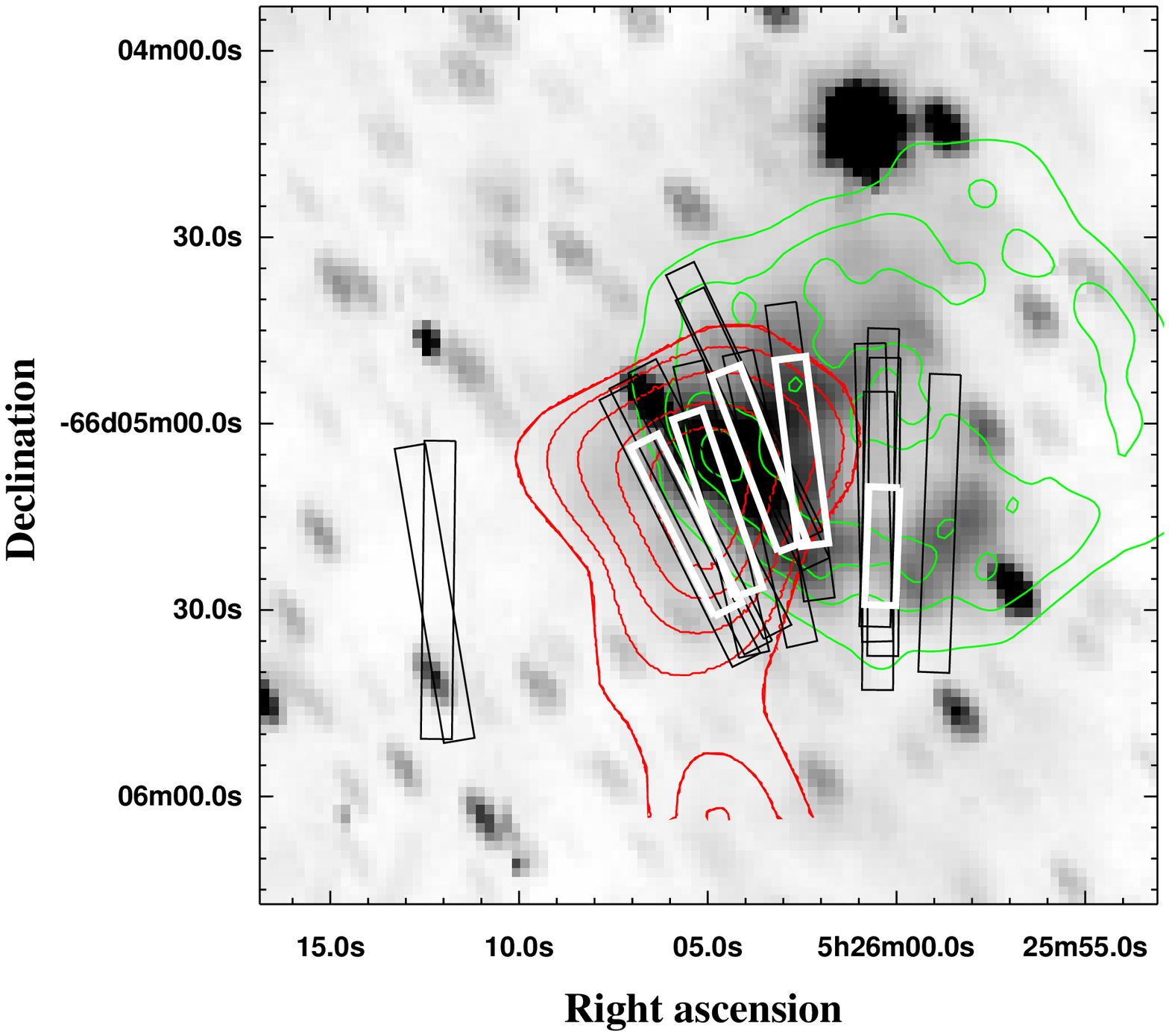}
\caption{Positions of the IRC NG/Ns slits superposed on the {\it AKARI}
N3 image of N49. Thirteen on-positions and two background positions 
are observed (black). We also mark where final averaged spectra are extracted (white). The five final spectra are referred to as ``P1'' to ``P5'' from east to west, respectively (see Fig. \ref{spec}). For comparison, the contours showing 6-cm radio continuum (green contours) and CO emission (red contours) are superposed. The radio data are from \citet{dick98}; the levels are at 2, 4, 6, 8, and 10 mJy beam$^{-1}$. The CO contours are taken from Figure 5 of \citet{ban97}; the levels are at 1.0, 1.3, 1.7, 2.0, 2.3, and 2.7 K km s$^{-1}$.  \label{fig1}}
\end{figure}

\clearpage

\begin{figure}
\epsscale{0.9}
\plotone{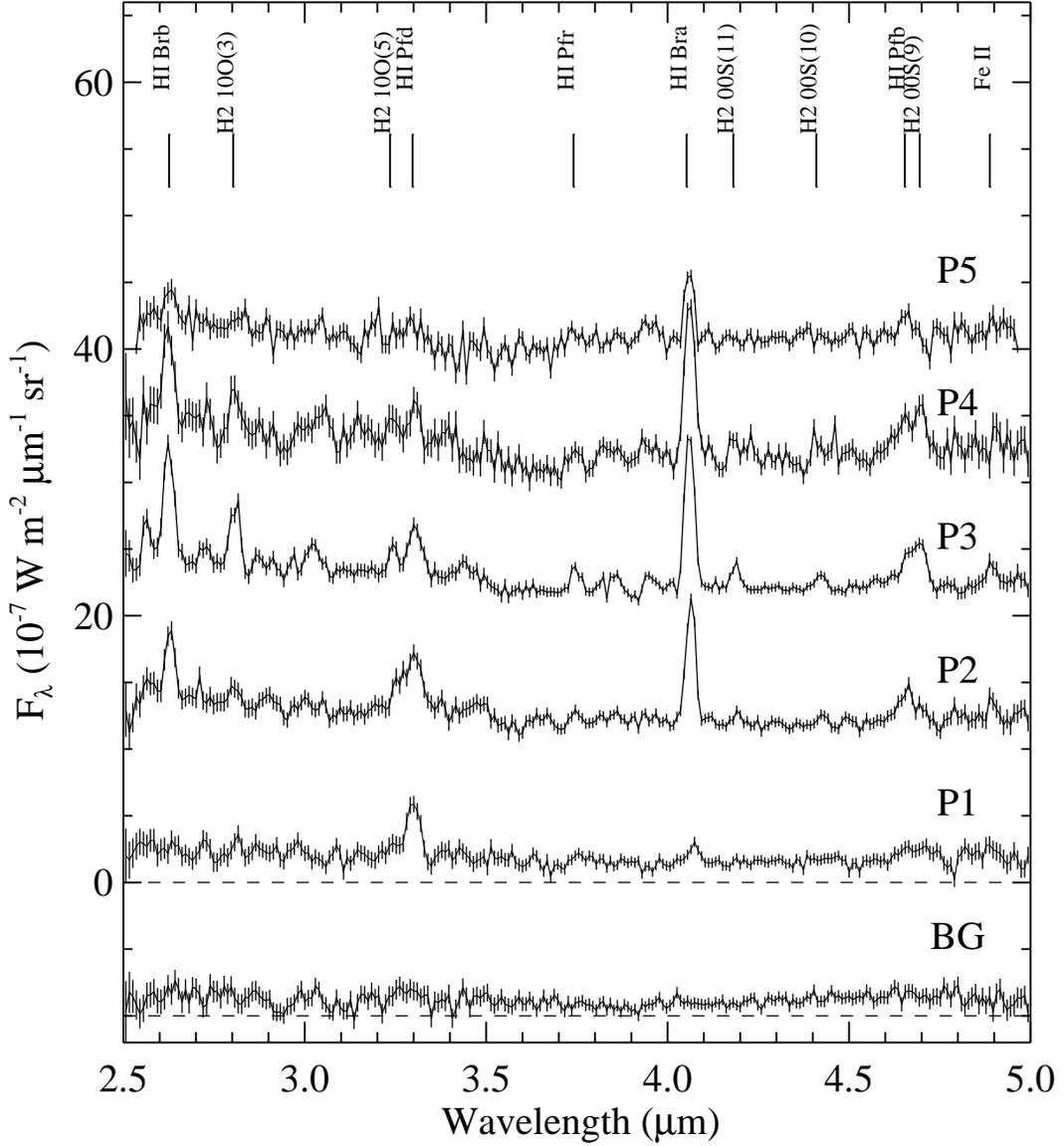}
\caption{Final spectra together with the background spectrum. For convenience, the spectra are shifted by 1, 2, 3, and 4 $\times10^{-6}$ W m$^{-2}$ \micron$^{-1}$ sr$^{-1}$ for P2 to P5, respectively. Background spectrum is shown in the bottom (shifted by -1 $\times10^{-6}$ W m$^{-2}$ \micron$^{-1}$ sr$^{-1}$). Noticeable lines are labeled at their vacuum wavelengths.  \label{spec}}
\end{figure}

\clearpage

\begin{figure}
\epsscale{1}
\plotone{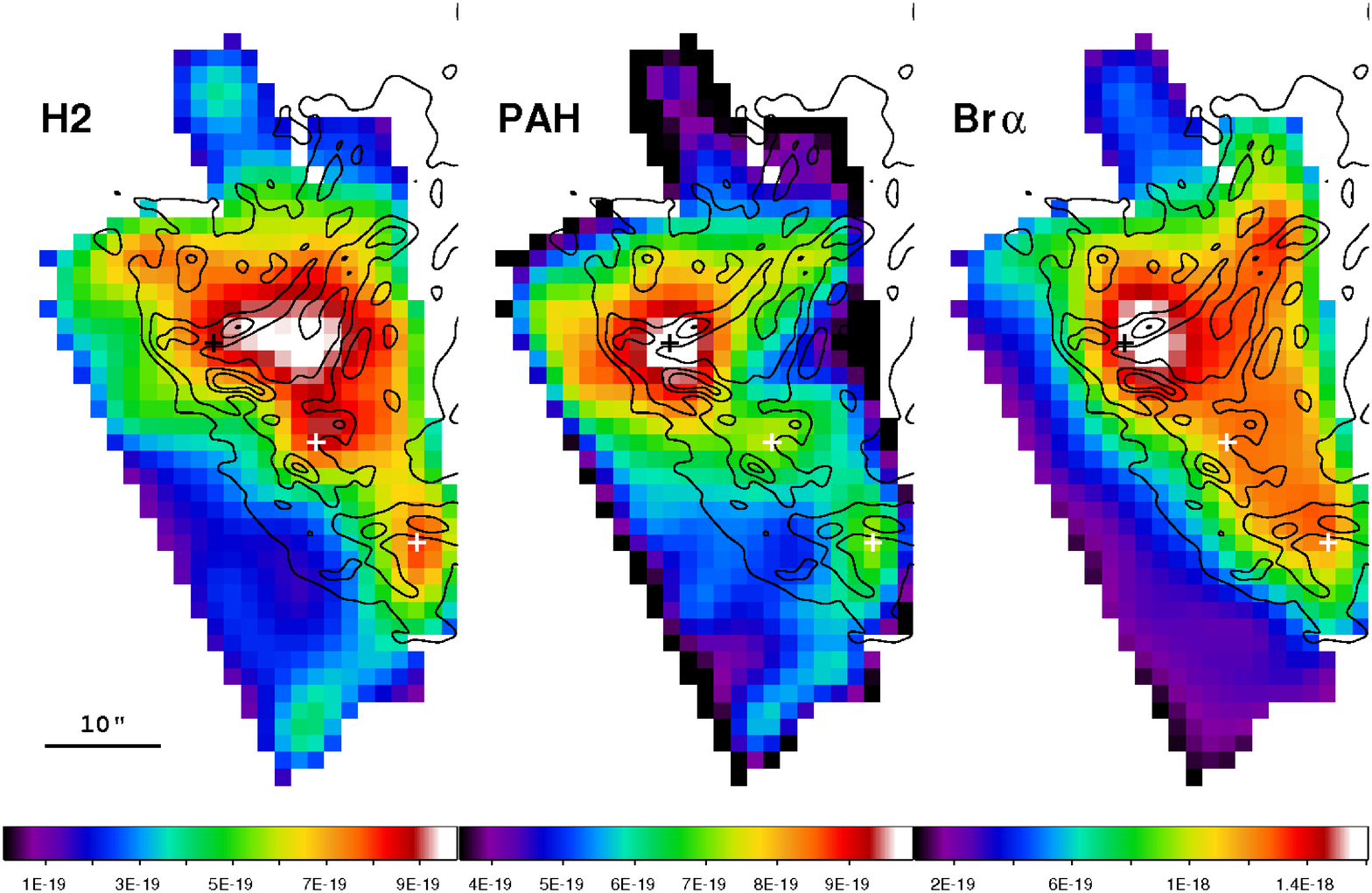}
\caption{H$_2$ 1-0 O(3), PAH, and Br$\alpha$ line maps of N49 by using $AKARI$ IRC spectra. For comparison, contours (black) are taken from an H$\alpha$ image of the $HST$ WFPC2 \citep{bil07}; these contours are 10, 30, 50, and 70 \% of the peak. The black cross marks the peak of PAH emission at ($\alpha_{2000}$, $\delta_{2000}$) = ($05^h26^m05.14^s, -66\degr05\arcmin03\arcsec$). The white crosses denote two clumps seen in the PAH line map. The scale bar represents 10\arcsec~or 2.4 pc at 50 kpc. The color bars are given in units of W m$^{-2}$ arcsec$^{-2}$. North is up, and east is to the left. \label{map}}
\end{figure}

\clearpage

\begin{figure}
\epsscale{0.7}
\plotone{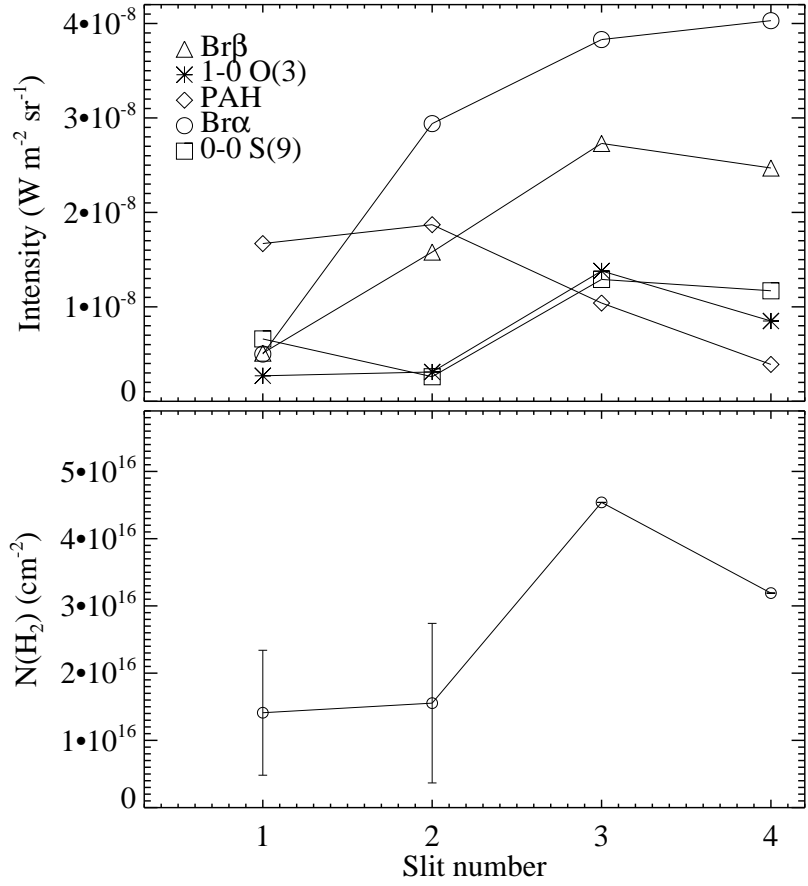}
\caption{Variation of line intensities and H$_2$ column densities with respect to the different positions (from P1 to P4). The estimation of the column densities are described in \S~3.2. P5 spectrum is excluded due to non-detection of any H$_2$ lines. \label{h2fit}}
\end{figure}
\clearpage

\begin{figure}
\epsscale{1}
\plotone{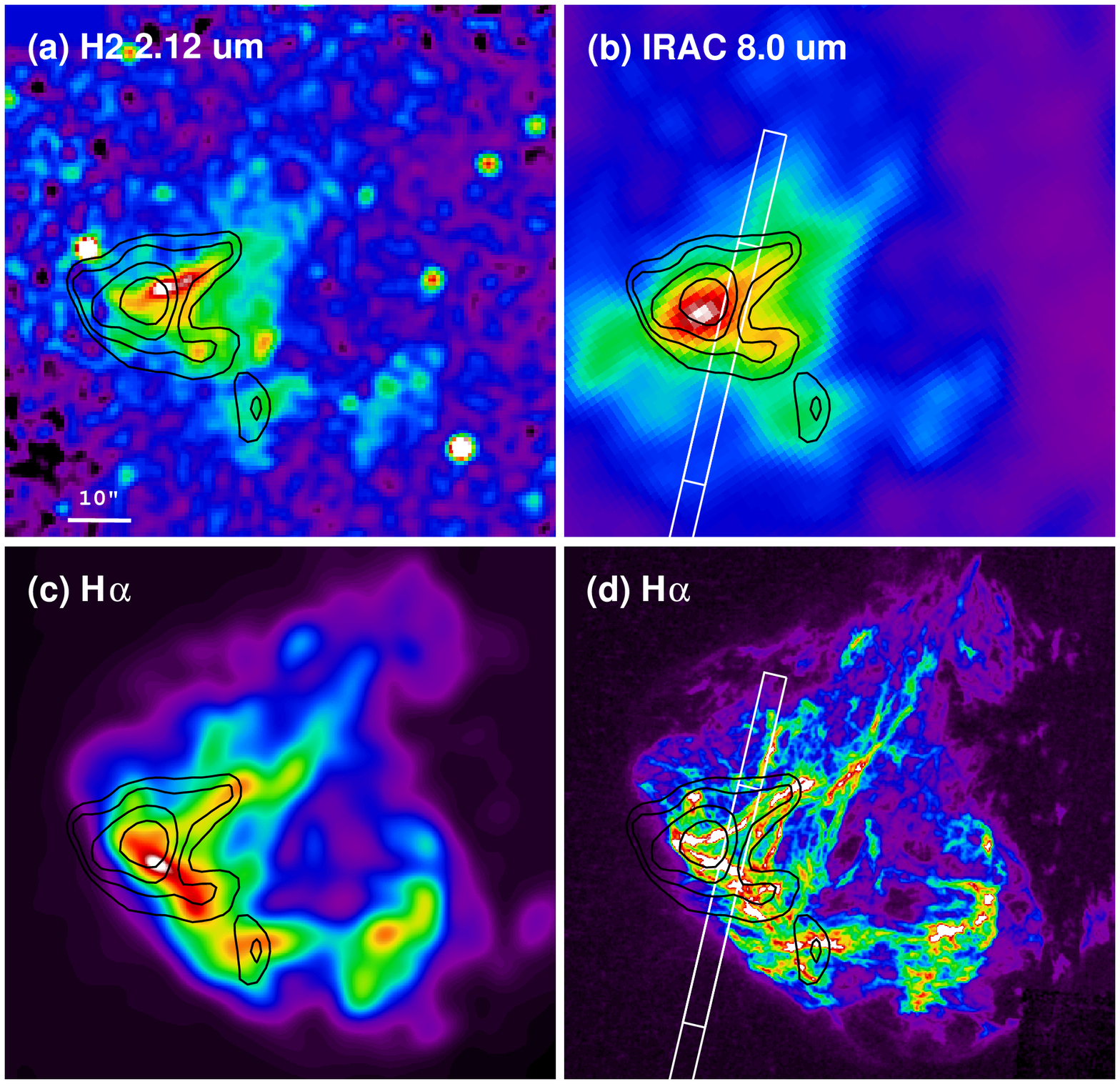}
\caption{Images of N49 in $(a)$ H$_2$ 2.12 \micron~\citep{dick95}, $(b)$ IRAC 8.0 \micron, $(c)$ and $(d)$ smoothed and original H$\alpha$ \citep{bil07}. $(c)$ H$\alpha$ image is convolved to the spatial resolution of $AKARI$ N3 image (FWHM $\sim 4.0 \arcsec$). Contours overlaid in all images are from the PAH 3.3 \micron~map of Figure \ref{map}, of which levels are at 6, 7, 8, and 9 $\times 10^{-19}$ W m$^{-2}$ arcsec$^{-2}$. North is up, and east is to the left. Positions of the $Spitzer$ IRS SL resolution slits are marked in $(b)$ and $(d)$. \label{aat}}
\end{figure}
\clearpage

\begin{figure}
\plotone{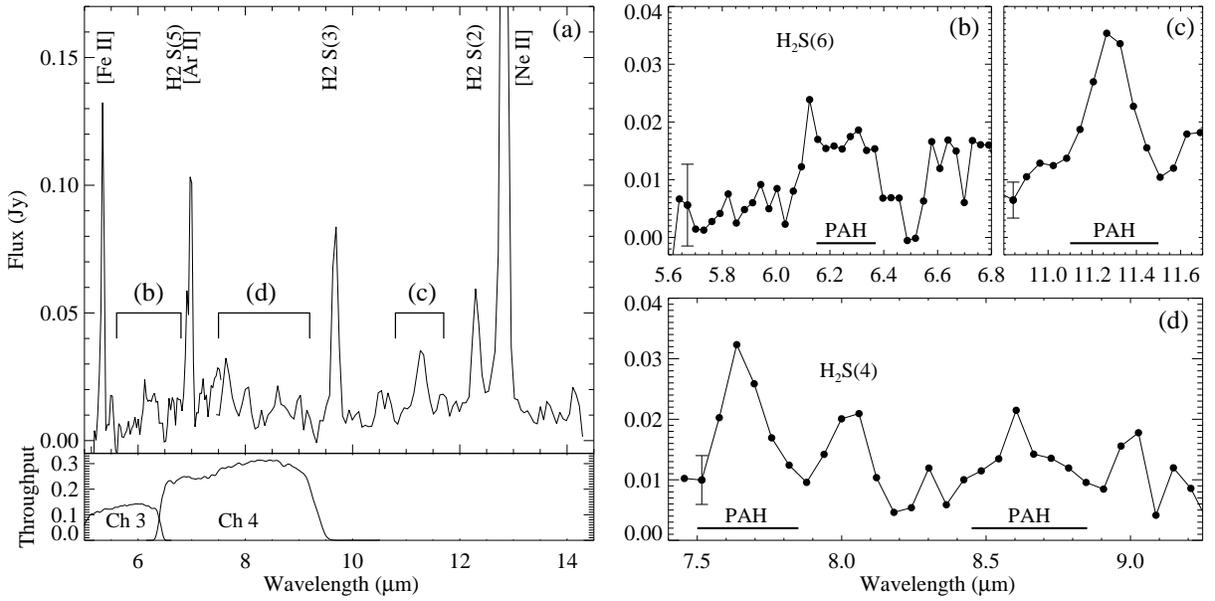}
\caption{$(a)$: IRS SL spectra of N49 with $Spitzer$ IRAC spectral response curves.   ``Nod 1'' position spectra are used only. The [Ne II] line at 12.8 \micron~is truncated for clarity. Dominant emission lines are labeled. $(b)-(d)$: Profiles of the PAH band emission features at 6.2, 11.3, and $7-9$ \micron~extracted from the IRS spectra. The extracted ranges are overplotted in $(a)$. Detected PAH bands with H$_2$ emission lines are also marked. Error bars at the left bottom of each panel represent the typical error for the spectra.  \label{irs}}
\end{figure}

\clearpage

\begin{figure}
\epsscale{0.7}
\plotone{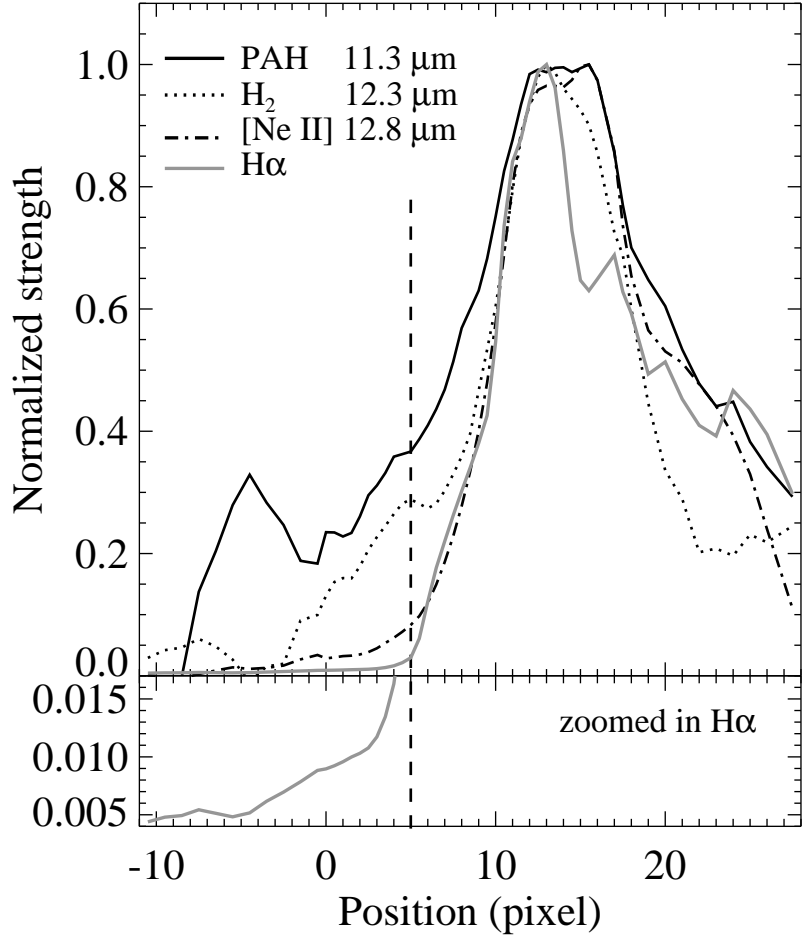}
\caption{$Top$: IRS SL line profiles of PAH $\lambda$ 11.3 \micron~(black solid), H$_2$ 0-0 S(2) $\lambda$ 12.3 \micron~(dotted), and [Ne II] $\lambda$ 12.8 \micron~(dash dot). For comparison, an one-dimensional cut of H$\alpha$ emission along the IRS slits (gray solid) is extracted from the $HST$ image smoothed to the pixel scale of the IRS. $Bottom$: The same profile of H$\alpha$, but zoomed in for the variation at lower levels. The dashed line represents the SNR boundary (the probable location of the shock front). One pixel corresponds to 1.8$\arcsec$, and the coordinate of the position for the zero pixel is ($\alpha$, $\delta$) = ($05^h26^m05^s, -66\degr05\arcmin28\arcsec$; J2000.0). The negative and the positive correspond to south and north, respectively. \label{prof}}
\end{figure}
\clearpage

\begin{figure}
\epsscale{0.9}
\plotone{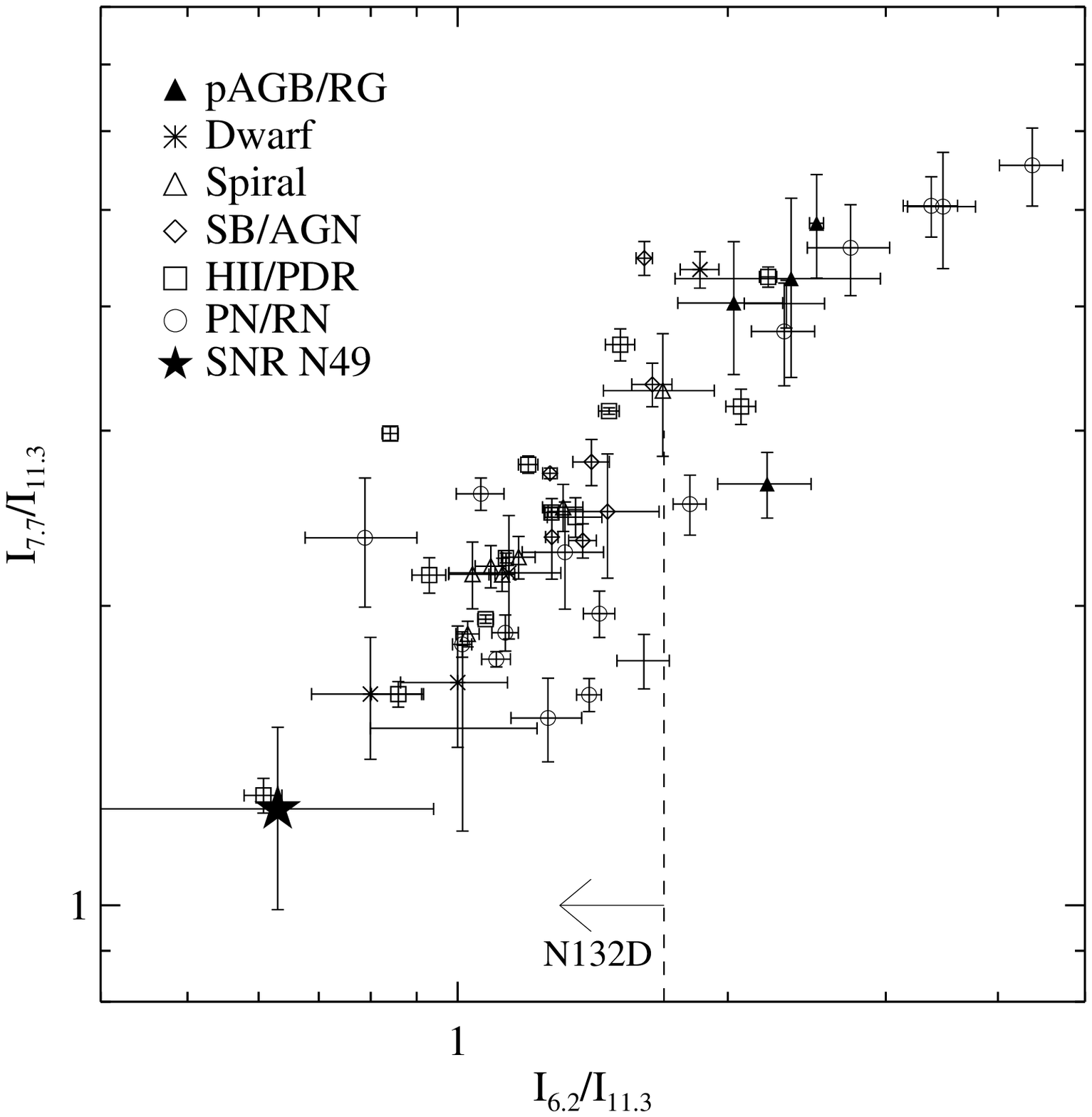}
\caption{PAH band ratios of SNR N49 as well as various objects from literatures \citep{verm02,sloan07,gal08,bsala09}. The ratio in N49 is derived from the $Spitzer$ IRS spectrum (\S~3.3). The upper limit of $I_{6.2}/I_{11.3}$ from the SNR N132D is marked as a dashed line \citep{tappe06}. Different symbols denote different types of objects. Few objects without any symbols are either Herbig Ae/Be or T Tauri stars or a Magellanic object. \label{ratio}}
\end{figure}
\clearpage

\begin{figure}
\plotone{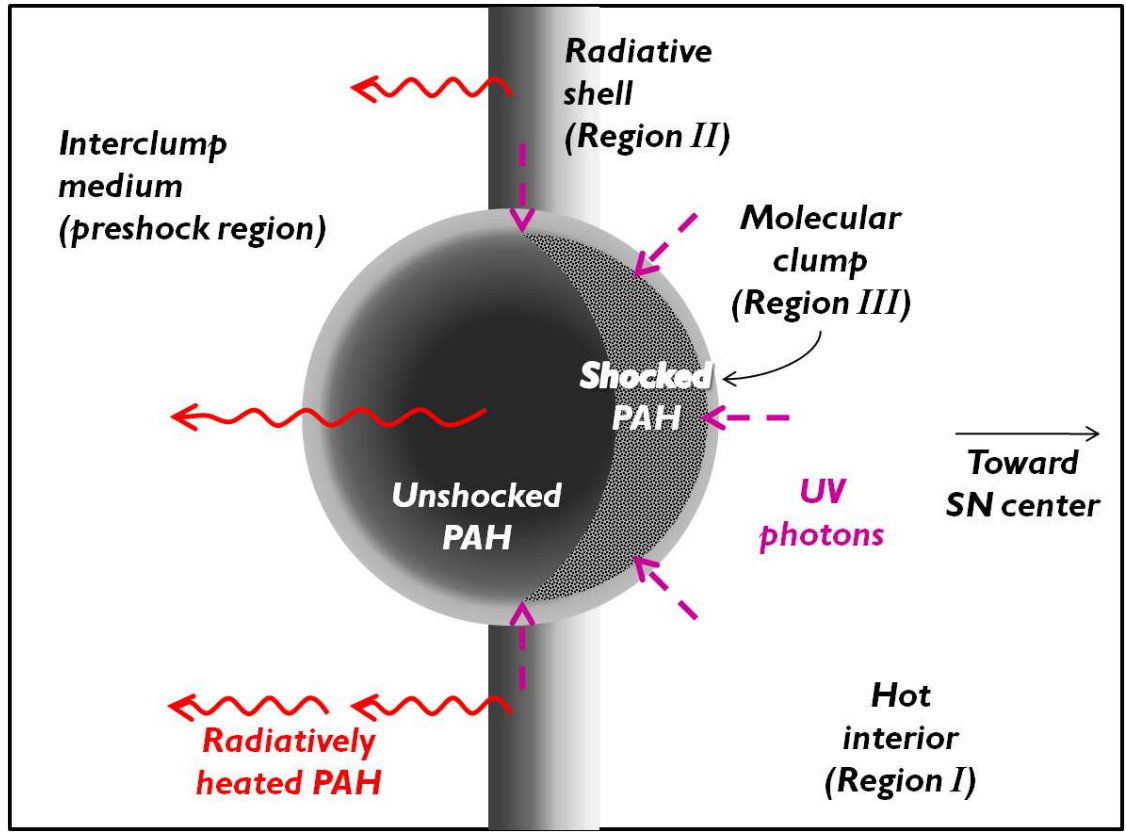}
\caption{Schematic diagram of local medium structures in N49. There are largely three kinds of medium; the hot interior, the radiative shell, and the molecular clumps emitting X-ray, ionic line emission, and molecular emission, respetively. Darker color indicates the higher survivability of PAH molecules. In the case of a dense molecular clump, its inner layer can remain unshocked. UV photons mainly originate from the radiative shell, more precisely, neighboring optical filaments around the molecular clump in N49. Survived PAHs associated either the radiative shell or the molecular clump can be heated by UV photons and produce detectable emission features.}\label{schem}
\end{figure}
\clearpage

\begin{deluxetable}{ccccccc}
\tablecaption{Summary of Observations \label{tbl-1}}
\tablewidth{0pt}
\tablehead{
\colhead{} & \colhead{} & \colhead{} & \colhead{} & \multicolumn{2}{c}{Field center}& \colhead{}\\
\cline{5-6}\colhead{Obs. ID} & \colhead{Sub. ID} & \colhead{Program\tablenotemark{a}} & \colhead{Date} & \colhead{R.A. (J2000)} & \colhead{Dec. (J2000)}  & \colhead{Note\tablenotemark{b}} 
}
\startdata

1420813 & 1 & ISMGN & 2008 Jun 12 & 05:26:12.1 &  -66:05:26.8 & BG \\
1420813 & 2 & ISMGN &2008 Jun 23 & 05:26:12.2 &  -66:05:27.3 & BG \\
1420925 & 1 & ISMGN &2008 Jul 10 & 05:26:04.9  &  -66:05:12.2 & P2 \\
1420926 & 1 & ISMGN &2008 Jul 08 & 05:26:03.8  &  -66:05:00.9 & P3 \\
1420926 & 2 & ISMGN &2008 Jul 08 & 05:26:04.0  &  -66:04:56.7 & P3 \\
1421925 & 1 & ISMGN &2008 Jul 10 & 05:26:05.7  &  -66:05:16.6 & P1 \\
1421925 & 2 & ISMGN &2008 Jul 10 & 05:26:05.5  &  -66:05:14.7 & P1 \\
1421926 & 1 & ISMGN &2008 Dec 20 & 05:26:02.6 &  -66:05:04.5 & P4 \\
1900451 & 1 & P3LMC &2008 Jun 25 & 05:26:04.6 &  -66:05:13.8 & P2  \\
1900452 & 1 & P3LMC &2008 Dec 25 & 05:26:03.4 &  -66:05:12.1 & P3  \\
1910233 & 1 & LMCNG &2009 Dec 11 & 05:25:58.9 &  -66:05:16.1 & N \\
1910233 & 2 & LMCNG &2009 Dec 12 & 05:25:56.8 &  -66:06:15.4 & N \\
1910233 & 3 & LMCNG &2009 Dec 12 & 05:26:00.6 &  -66:05:08.7 & P5 \\
1910234 & 1 & LMCNG &2009 Dec 13 & 05:26:00.3 &  -66:05:13.4 & P5 \\
1910234 & 2 & LMCNG &2009 Dec 13 & 05:26:00.5 &  -66:05:18.9 & P5\\
1910234 & 3 & LMCNG &2009 Dec 15 & 05:26:00.6 &  -66:05:11.1 & P5 \\

\enddata
\tablecomments{}
\tablenotetext{a}{Name of the $AKARI$ mission programs: ``ISM in our Galaxy and nearby galaxies'' (ISMGN, PI: H. Kaneda), ``Observations of the Magellanic Cloud'' (P3LMC, PI: T. Onaka), and ''Near infrared spectroscopic observations of red objects in the Large Magellanic Cloud'' (LMCNG, PI: T. Onaka). }
\tablenotetext{b}{Classification of the final spectra after averaging separate spectra extracted from overlapping region. Spectra with the same note are used for one final spectrum. BG: background spectrum, P1--P5: source spectra extracted from five different regions, respectively (see Fig. \ref{fig1}). N: spectrum not used for the final spectra owing to either a single observation under a poor detector condition or a wrong pointing.}
\end{deluxetable}

\clearpage

\begin{deluxetable}{ccccccc}
\tablecolumns{7}
\tablecaption{Detected lines and their intensities from AKARI IRC Data for N49 \label{tbl-2}}
\tablewidth{0pt}
\tablehead{
\colhead{} & \colhead{} & \multicolumn{5}{c}{Intensity ($10^{-8}$ W m$^{-2}$ sr$^{-1}$)} \\
 \cline{3-7} \colhead{Line ID} & \colhead{$\lambda_{ref}\tablenotemark{a}$} & \colhead{P1} & \colhead{P2} & \colhead{P3} & \colhead{P4} & \colhead{P5} 
}
\startdata
Br$\beta$  & 2.626 & $\leq$0.34 & 1.58$\pm$0.30 
& 2.73$\pm$0.27 & 2.47$\pm$0.53 & 0.77$\pm$0.34 \\
H$_2$ 1-0 O(3) & 2.803 & 0.27$\pm$0.15\tablenotemark{b} & 0.31$\pm$0.25 
& 1.38$\pm$0.23 & 0.85$\pm$0.40 & $\leq$0.47 \\
H$_2$ 1-0 O(5) & 3.235 & 0.32$\pm$0.15\tablenotemark{b} & 0.75$\pm$0.38 
& 0.58$\pm$0.12\tablenotemark{b} & 0.42$\pm$0.24\tablenotemark{b} & $\leq$0.57 \\
Pf$\delta$+PAH\tablenotemark{c} & 3.297 & 1.72$\pm$0.31\tablenotemark{d} 
& 2.16$\pm$0.51 & 1.42$\pm$0.25\tablenotemark{d} & 0.79$\pm$0.37\tablenotemark{d} & $\leq$0.73   \\
PAH\tablenotemark{e} & 3.300 & 1.67$\pm$0.35 &1.87$\pm$0.53 
& 1.04$\pm$0.28 & 0.39$\pm$0.45 & ---\\
PAH\tablenotemark{f} & 3.400 & --- & 0.61$\pm$0.36\tablenotemark{f} &0.80$\pm$0.34\tablenotemark{f} &--- &--- \\
Pf$\gamma$ & 3.741 & $\leq$0.24 & 0.22$\pm$0.09\tablenotemark{b} 
& 0.51$\pm$0.13 & $\leq$0.44  & $\leq$0.45\\
Br$\alpha$ & 4.052 & 0.50$\pm$0.16 & 2.94$\pm$0.15  
& 3.83$\pm$0.13 & 4.03$\pm$0.25 & 1.52$\pm$0.17\\
H$_2$ 0-0 S(11) & 4.181 & $\leq$0.09 & 0.23$\pm$0.08\tablenotemark{b} 
& 0.50$\pm$0.11 & 0.42$\pm$0.14\tablenotemark{b}  & $\leq$0.48  \\
H$_2$ 0-0 S(10) & 4.410 & $\leq$0.20 & 0.20$\pm$0.09\tablenotemark{b}  
& 0.35$\pm$0.14 & $\leq$0.51 & $\leq$0.29 \\
Pf$\beta$ & 4.654 & $\leq$0.57  & 0.84$\pm$0.11\tablenotemark{b} 
& 0.79$\pm$0.11\tablenotemark{b} & 0.88$\pm$0.17\tablenotemark{b} & 0.64$\pm$0.24\\
H$_2$ 0-0 S(9)  & 4.695 &$\leq$0.44& 0.26$\pm$0.20\tablenotemark{d}  
& 1.29$\pm$0.23\tablenotemark{d} & 1.17$\pm$0.35\tablenotemark{d} & $\leq$0.55\\
$$[Fe \textsc{ii}] & 4.889 & $\leq$0.35 & 0.40$\pm$0.16\tablenotemark{b}  
& 0.53$\pm$0.23 & $\leq$0.48 & 0.49$\pm$0.18\tablenotemark{b}\\
\hline
& & \multicolumn{5}{c}{Ratio} \\
\cline{3-7}&Case B\tablenotemark{g}&P1&P2&P3&P4&P5\\
\hline
Br$\beta$/Br$\alpha$ & 0.60& $\leq$0.68 & 0.54$\pm$0.11 & 0.71$\pm$0.07 & 0.61$\pm$0.14 & 0.51$\pm$0.23\\ 
Pf$\delta$\tablenotemark{c}/Br$\alpha$ & 0.10 & 3.44$\pm$1.26 & 0.73$\pm$0.18 & 0.37$\pm$0.07 & 0.20$\pm$0.09 & $\leq$0.48\\
Pf$\gamma$/Br$\alpha$ &0.135& $\leq$0.48 & 0.07$\pm$0.03& 0.13$\pm$0.03& $\leq$0.11 & $\leq$0.30\\
Pf$\beta$/Br$\alpha$ &0.20& $\leq$0.57& 0.29$\pm$0.04 & 0.21$\pm$0.03 & 0.22$\pm$0.04 & 0.42$\pm$0.16 \\

\enddata
\tablecomments{Intensities with errors of detected lines are given. Upper limits are at the 2$\sigma$ level. }
\tablenotetext{a}{Vacuum wavelength of a line in \micron.}
\tablenotetext{b}{FWHM fixed to the instrumental width ($\sim0.03$ \micron) for a gaussian fit.}
\tablenotetext{c}{Total intensity of the feature at 3.3 \micron, which can be mainly contributed by either 3.3 \micron~PAH or Pf$\delta$ emission.}
\tablenotetext{d}{Central wavelength fixed for a gaussian fit.}
\tablenotetext{e}{Estimation of PAH intensity by subtracting the total intensity of the 3.3 \micron~feature by the scaled Br$\alpha$ intensity (i.e. 10\% of the Br$\alpha$ intensity).}
\tablenotetext{f}{Intensity of a minor PAH feature at 3.4 \micron. Because of the curvature around the feature, the baseline is fitted by using a second order polynomial. Due to the ambiguous identification, the measured intensities can be regarded as an upper limit of the 3.4 \micron~feature intensity.}
\tablenotetext{g}{Theoretical recombination line ratio in the case of ``On-the-spot'' approximation (so called ``Case B'') with $n_e\sim10^2-10^7$ cm$^{-3}$ and $T_e\sim5-30\times10^3$ K \citep{hum87}.}
\end{deluxetable}

\end{document}